\documentclass[twocolumn]{aastex6}

\begin{document}


\title{Variable stars with the \textit{Kepler} space telescope}

%% Use \author, \affil, plus the \and command to format author and affiliation 
%% information.  If done correctly the peer review system will be able to
%% automatically put the author and affiliation information from the manuscript
%% and save the corresponding author the trouble of entering it by hand.
%%
%% The \affil should be used to document primary affiliations and the
%% \altaffil should be used for secondary affiliations, titles, or email.

%% Authors with the same affiliation can be grouped in a single
\author{L\'aszl\'o Moln\'ar\altaffilmark{1}, R\'obert Szab\'o\altaffilmark{1}, Emese Plachy\altaffilmark{1}}
\affil{Konkoly Observatory, Research Centre for Astronomy and Earth Sciences, Konkoly Thege Mikl\'os \'ut 15-17, H-1121 Budapest, Hungary}


\altaffiltext{1}{molnar.laszlo@csfk.mta.hu}

\begin{abstract}
The \textit{Kepler} space telescope has revolutionised our knowledge about exoplanets and stars and is continuing to do so in the K2 mission. The exquisite photometric precision, together with the long, uninterrupted observations opened up a new way to investigate the structure and evolution of stars. Asteroseismology, the study of stellar oscillations, allowed us to investigate solar-like stars and to peer into the insides of red giants and massive stars. But many discoveries have been made about classical variable stars too, ranging from pulsators like Cepheids and RR Lyraes to eclipsing binary stars and cataclysmic variables, and even supernovae. In this review, which is far from an exhaustive summary of all results obtained with \textit{Kepler}, we collected some of the most interesting discoveries, and ponder on the role for amateur observers in this golden era of stellar astrophysics. 
\end{abstract}




\section{Introduction} 
The \textit{Kepler} space telescope, as a Discovery-class space mission, was built to carry out a specific set of tasks to meet well-defined goals. It was conceived to do exoplanet statistics, and determine $\eta_{Earth}$, the frequency of small, rocky planets within the habitable zone of their stars \citep{Borucki-2016}. However, it turned out to be, as many astronomers had hoped, much more than just an exoplanet-statistics mission. It is fair to say that during the last few years, \textit{Kepler} has not only transformed our understanding of exoplanets but also revolutionized the field of stellar astrophysics. 

But all good things come to an end, and the primary mission of \textit{Kepler} ended in 2013, after collecting data from more than 160 thousand stars in the same patch of sky for four years, quasi-continuously. This was not the end of the telescope itself though. With only two functioning reaction wheels remaining to point the spacecraft, an ingenious new mission called K2 was initiated \citep{Howell-2014}. The telescope that was once built for a singular purpose, was transformed into a community resource, open to any targets available within its new observing fields. It observes in 80-day campaigns along the Ecliptic, and at the time of writing finishes its 10th observing run with only minor technical problems. 

The discoveries of the primary and extended missions of \textit{Kepler} can already fill books. \textit{Kepler}, along with the other space photometric missions, opened up a new window for us to explore what was considered utterly unreachable a century ago: the insides of stars. In this review we focus on the most important or interesting results about variable stars: stars that show light variations due to excited pulsation modes, turbulent convection,  binarity, cataclysmic or eruptive events. Some of these are out of reach of an amateur astronomer, but most of them are interesting to all variable star enthusiasts. 

\section{Kepler data}

\begin{figure*}
\includegraphics[width=1.0\textwidth]{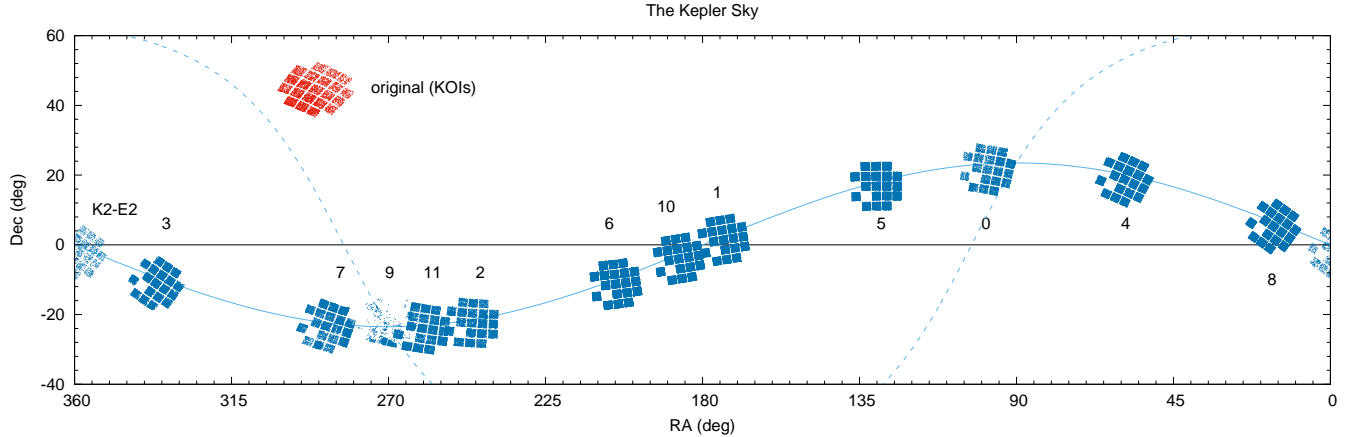}
\caption{The observing fields of \textit{Kepler} in the sky. For the original field (red) we only plotted the KOIs (\textit{Kepler} Object of Interest) instead of all targets. K2 fields include the positions of all observed targets and pixel mosaics. In Campaign 9 most of the pixels were assigned ot a smaller, but continuous area to search for microlensing events, hence the sparse coverage for the other CCD modules. }
\label{fig:k2sky}
\end{figure*}

The success of  \textit{Kepler} mission resulted from the combination of its unprecedented photometric accuracy ($10^{-5}-10^{-6}$ relative precision), the length and the continuity of the observations, as well as the fast data sampling (1 and 30 minute cadence), that led to discoveries of light variation well below  millimagnitude level and insight into the details of long-term behavior of a large number of stellar targets. 

However, like all instruments, \textit{Kepler} too has unwanted artificial effects that contaminate the beautiful light curves. The most puzzling issue for which no perfect solution exists, is stitching the individual data quarters to each other. In order to keep the solar panels pointed towards the Sun, the telescope had to roll 90 degrees after every three month, and as a consequence the targets ended up on different detectors for every quarter of a year, often causing significant differences in the measured flux. The correction of these differences requires scaling, shifting, and detrending of the observed flux, and are especially challenging for stars that show slow and irregular variability. \textit{Kepler} light curves are also affected by a sinusoidal variability with the \textit{Kepler}-year (372.5 d, the orbital period of spacecraft around the Sun) due to the change of the thermal properties of the telescope elements. The amplitude and the phase of this effect is dependent on the position of the star within the field of view \citep{Banyai-kiss-2013}.

Astronomers do not have to bother with these problems in the K2 mission any more---but other issues appeared instead. Due to the inherent instability of the positioning with only two functioning reaction wheels, the attitude of the telescope drifts (it rolls back and forth about the optical axis) and corrective manoeuvres are required at every 6 hours. The roll and correction causes stars to move across slightly differently sensitive pixels, causing distinctive 6-hour jumps to be present in the light curves. These effects are the strongest for stars that fall close to the edges of the field of view. Another source of noise is the zodiacal light, light scattered from the fine dust in the inner Solar System, that increases the background noise towards the end of each campaign. 

Nevertheless, thanks to the Earth-trailing, i.e., heliocentric orbit of \textit{Kepler} the data are devoid of other problems typical of space telescopes operating in low-Earth orbit. Instruments like CoRoT and the BRITE satellites are prone to scattered light from the Earth and the Moon, temperature changes when crossing the shadow of the Earth, and degradation or gaps in the data when they cross the swarm of charged particles, called South-Atlantic Anomaly. The patches of sky observed or targeted by \textit{Kepler} are shown in Fig.~\ref{fig:k2sky}. 

The core exoplanet science of the original mission was led by the \textit{Kepler} Science Team. But in order to exploit the wealth of stellar oscillation data provided by \textit{Kepler} optimally, a large collaboration (Kepler Asteroseismic Science Consortium) was also formed, which consists of some 600 scientists around the globe, and produced most of the results presented in this paper.

\section{Solar-like oscillations}
Probably the single greatest breakthrough that \textit{Kepler} delivered for stellar astrophysics is the huge number of stars where solar-like oscillations have been detected. Detailed asteroseismic analysis is now routinely used to determine the physical parameters of stars, and, by extension, exoplanets. Asteroseismic modeling, e.g., fitting the spectrum of observed oscillation modes with values calculated from theoretical models, is extremely powerful, for multiple reasons. First, it requires photometric measurements instead of spectroscopic ones, so it can be done for a large number of stars simultaneously, with a relatively small telescope. (Although with the need of very high precision: oscillation signals are closer to the $\mu$mag level than the mmag level usually accessible with ground-based instruments.) Secondly, although resolved observations are not possible for distant stars, detailed seismic (i.e. intensity) observations allow for the determination of global parameters, like mass, radius, and age much more precisely, than with any other methods. The typical precision is 3-5\% in mass and radius and 10\% in age for main-sequence stars. The latter can be appreciated if we mention that the age of a typical (not so young) main sequence star can be determined with rather large (30-50\%) uncertainties based on spectroscopic information alone. And any age or radius information about exoplanets is only as good as our knowledge on their host stars. 

Before CoRoT and \textit{Kepler}, seismic information was available only for a handful of stars, mainly through spectroscopic observations or small space telescopes, like WIRE. In contrast, \textit{Kepler} delivered seismic information for hundreds of main-sequence stars (brightest ones), and over 15 thousand red giants in the original \textit{Kepler} field. Many more are expected from the ongoing K2 Mission. This amount of stars with accurate asteroseismic information already makes galactic archaeology and population studies possible. Thus, we can learn more about the history of our Galaxy, the stellar formation rate, initial mass function, etc., especially in conjunction of Gaia, the flagship mission of the European Space Agency, which provides accurate distances and proper motions for roughly one billion stars. 

Kepler-16 A\&B, the two solar-like components of a visual binary allowed the execution of an exquisite proof--of--concept seismological study \citep{Metcalfe-2012}. The masses of these stars are 1.10 $\pm$ 0.01 and 1.06 $\pm$ 0.01 solar masses and were resolvable even with \textit{Kepler}'s not-so-good resolving power (it was optimized to gather as much photons, as possible, hence the large pixels and relatively low spatial resolution). Thus, the oscillation spectra were derived for both stars, and their parameters could be determined independently and without any prior assumptions. Reassuringly, the age of the two objects (6.8 $\pm$ 0.4 billion years) and their bulk chemical compositions were found to be identical. That is what one would expect if the two objects were formed from the same blob of interstellar material at the same time. What's more, 16 Cyg A and B are close enough to us that their radii can be measured directly, and as such, they can help us validate the asteroseismic relations that are based on our knowledge of the Sun. Data from the CHARA (Center for High Angular Resolution Astronomy) Array revealed that the seismic and interferometric radii of solar-like stars agree within a few percent \citep{White-2013}.  

Not only individual stars can be scrutinized by applying asteroseismological methods. Models can be constrained even more if our object belongs to a stellar cluster, since in this case the cluster members were presumably formed from the same molecular cloud roughly at the same time (although recently evidence is accumulating for multi-epoch star formation in globular clusters). In the original \textit{Kepler} field there were four open clusters (NGC\,6811, NGC\,6819, the very old NGC 6791, and NGC 6866, the youngest of the four). Their red giant population was investigated in great detail in a large number of studies based on the \textit{Kepler} data. For example, a new methodology was developed to determine cluster membership probability based on seismic constraints \citep{Stello-2011} and the very uncertain mass loss was studied along the red giant branch evolution \citep{Miglio-2012}, again, based on \textit{Kepler}'s seismic information. 

The detection of seismic signals from the main sequence of an open cluster is a much harder task, since main-sequence stars are usually much fainter then their evolved fellow cluster members. Therefore this breakthrough was only recently reached: \citet{Lund-2016} succeeded in detecting seismic signals from main-sequence stars in the Hyades cluster for the first time. Many more, already well studied, fiducial clusters will be observed during the K2 extension including the Pleiades, M44, M67, M35, etc., that will allow a stringent test for our knowledge of stellar evolution. \citet{Miglio-2016} detected solar-like oscillation in the K giants belonging to the globular cluster M4 in the K2 Mission. This is also remarkable given the difficulty presented by the crowded nature of the target.

To close this section we briefly mention the synergies of stellar oscillatons with exoplanetary science. The first one is the measurement of the radius of the previously discovered hot Jupiter, HAT-P-7b in the \textit{Kepler} field. With the help of asteroseismology the radius of the host star was measured by more than an order of magnitude more precisely compared to traditional methods \citep{Christensen-Dalsgaard-2010}. 
Similar improvement was achieved regarding the radius of the exoplanet, since the transit depth to the first order scales with the ratio of the planetary and stellar radii.

The second nice example for the complementary nature of exoplanetary and stellar astrophysics results is the Kepler-444 planetary system \citep{Campante-2015}. The host star is a red dwarf with a mass of  0.758 $\pm$ 0.043 solar masses and can be found in constellation Lyra at a distance of 117 light years from the Solar System. There are five rocky planets orbiting the host star as revealed from from the \textit{Kepler} photometry, since all of them show transits. The most important aspect of the system is its age. From seismology it was found to be 11.2 $\pm$ 0.4 Gyr. First, it is astonishing, how accurate this age determination is. Secondly, this system is more than twice as old as our Solar System, and had to be formed right after the formation of the Milky Way galaxy. Another consequence is that the formation of planetary systems is not restricted to later, more metal-rich stars. Although the planets in the Kepler-444 system are too close to their star to be habitable, it seems that at least in some systems there has been ample time for undisturbed biological evolution -- if other conditions are also right and conducive. Again, this result also relies on the investigation of stellar oscillations in the form of tiny intensity variations in the red dwarf.

\section{From red giant variables to Miras}
Oscillations in stars come in two distinct flavors. The \textit{p}-modes, or sound waves, are governed by forces exerted by gas pressure, and are usually the strongest in the envelope, close to the surface. The other bunch is the \textit{g}- or gravity modes. These are like waves on a pond, but in solar-like stars they are locked away in the core, and, as of yet, unobservable by us. But as stars leave the Main Sequence, their cores begin to contract, and their envelopes start to expand. Frequency regimes of the two types of modes, nicely separated on the Main Sequence, begin to shift towards each other, as they depend from the density of the core and envelope, respectively. Soon the frequencies of some \textit{p-} and \textit{g}-modes start to overlap, and so-called mixed modes appear: these travel in the core mostly as \textit{g}-mode, but may traverse the steep core-envelope density boundary, and continue outwards more like a \textit{p}-mode. It was soon realized that mixed modes may carry extremely valuable information about the insides of red giants, down to the very cores they come from. By an interesting twist of nature, we can now investigate the centers of these stars while that of the Sun remains unreachable for us. 

For example it was possible to derive the rotation rate of the red giant envelope and the core independently -- the rotation frequency was 
found to be at least ten times greater for the core \citep{Beck-2012}. This is actually what one would expect if the core contracts and the envelope expands -- it is enough to think about a ballerina or a skater performing a pirouette with stretched or pulled in arms. All these phenomena obey the same physical principle, namely the conservation of angular momentum. \textit{Kepler} was able to probe the core rotation of many red giants, and suggested that as stars get older, their cores spin down \citep{Mosser-2012}. This, too, was expected, as white dwarfs, the remnants of the cores of now gone stars, rotate slowly. However, the actual physical process(es) that transport angular momentum away from the core into the envelope are still hotly debated. 

\begin{figure*}
\includegraphics[width=1.0\textwidth]{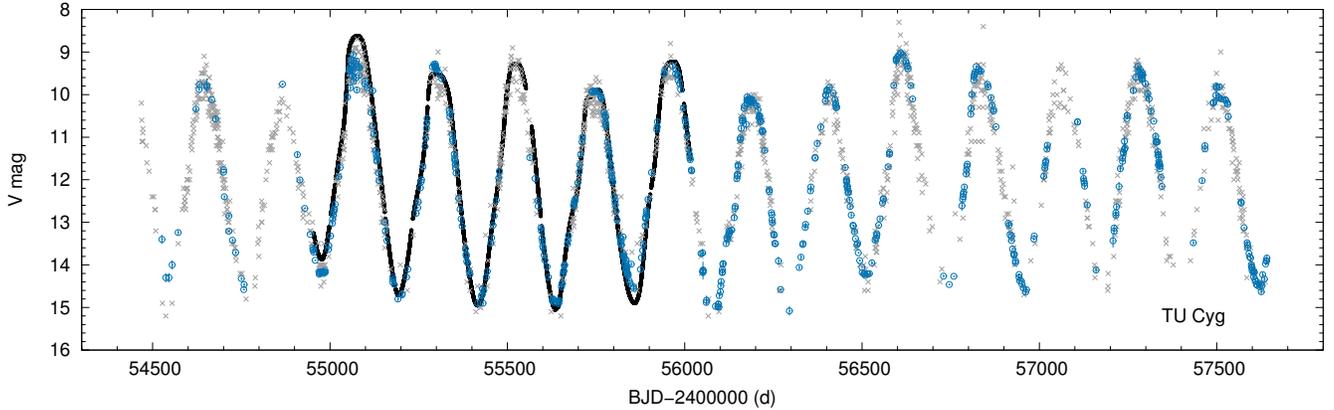}
\caption{A comparison of the AAVSO visual (grey), digital \textit{V}-band and green-filter (blue) observations and the \textit{Kepler} light curve (black) of the Mira star TU\,Cyg. Here we used the data from between 01/01/2009 and 16/09/2016. The corrected \textit{Kepler} data, published by \citet{Banyai-2013}, have is scaled to the amateur observations.  }
\label{fig:tucyg}
\end{figure*}

Another remarkable achievement was reached by the members of the \textit{Kepler} Asteroseismic Science Consortium. Based on the investigation of period spacing between gravity modes it is possible to distinguish between H-shell burning and He-core burning stars \citep{Bedding-2011}, i.e. different evolutionary states -- information that is otherwise hidden deep inside these stars, and is not reflected in their surface or global parameters. Yet another major breakthrough in stellar astrophysics thanks to \textit{Kepler} and its unbeatable photometric precision. 

The concept that with seismology we can in fact map the stellar interior can be demonstrated with another recent result. \textit{Kepler} observed over 15 thousand red giants and almost all of them show solar-like oscillations. But 10-15\% of them are peculiar: the amplitude of some of the nonradial (non-axysymmetric, namely $\ell=2$ modes) are significantly suppressed, these can be hardly seen in the oscillation spectrum. What causes this dichotomy between the normal amplitude and suppressed amplitude red giants? \citet{Fuller-2015} had an idea: {\it magnetic greenhouse effect}. It can be shown that if a strong magnetic field is present in the stellar interior then it can reflect and disperse certain incoming waves, that will be missing from the frequency spectrum. The critical magnetic field has a minimum in the H-burning shell that surrounds the He-core in these stars, therefore we can infer the minimum magnetic field at this specific point, which turns out to be $10^5-10^7$ Gauss, much stronger than previously thought, but still small to significantly alter the stellar structure. Voila: an ingenious way to see what is "hidden behind substantial barriers" -- to quote Sir Arthur Eddington, the famous astrophysicist \citep{Eddington-1926}.

Still, although solar-like oscillations are now the Swiss Army knives for many professional astronomers, their observations are mostly out of reach for amateur astronomers. But as we go to larger and larger red giants, the nice and tidy relation between the dominant periods and amplitudes of the oscillations starts to break down. At periods above about 10 days, the observed amplitudes start to increase much faster than below \citep{Banyai-2013}. This could be the transition from pure stochastic oscillations towards more coherent pulsations that lead to the more familiar landscape of semiregular and Mira stars above periods of 100 days. One Mira, TU\,Cyg, is shown in Fig.~\ref{fig:tucyg}, together with the more extended AAVSO observations. 

\section{Cepheids}
One century ago, the analysis of Cepheid stars revealed that pulsating stars exist, and that we can measure extragalactic distances with them. Since then, they have been under intense scrutiny, from all aspects. However, Cepheids are relatively rare stars in the Milky Way, and a thorough search for targets in the original \textit{Kepler} field came up with a single classical Cepheid star, V1154 Cyg \citep{Szabo-2011}. The first analysis and a later study based on 600 days of observations both showed that V1154 Cyg is a well-behaved Cepheid, e.g., it pulsates in the fundamental mode, without any signs of additional modes down to the sub-millimagnitude level. It was far from boring though: the quasi-continuous observations revealed that the variation of Cepheids may not be strictly periodic. Individual pulsation cycles of V1154 Cyg showed small changes in their respective shapes. Fluctuations up to 20-30 minutes occur in the O--C timings, but these average out on a time scale of $\approx$ 15 cycles, and the 4.9-d pulsation period remains stable \citep{Derekas-2012}. The physical cause behind this phenomenon is, so far, unclear: one hypothesis is that emerging convective hot spots can be responsible for the observed pulsation jitter \citep{Neilson-2014}. 

The latest analysis, based on the entire four-year data set, led to further discoveries. The star shows a 159--day long modulation cycle, reminiscent of a very low-level Blazhko effect . The \textit{Kepler} data was accurate enough to tease out the signals of granulation noise hiding under the large pulsation. This is the first firm detection of granulation in a Cepheid star: the effective timescale of the granulation cells is about 3.5 d that agrees with the scaling relations derived for smaller red giants \citep{derekas-2016}. But there was one thing that \textit{Kepler} would have been able to detect comfortably, but did not: solar-like oscillations. Apparently, coherent, large-scale pulsation inhibits the excitation of other, evanescent waves. 

\textit{Kepler} measured DF Cyg, a 24.9--day period type-II Cepheid as well. This star belongs to the RVb subclass of RV Tau variables that exhibit slow, large-scale mean brightness variations on top of the pulsation. During the observations a sudden period change was detected that was accompanied with an increase of the pulsation amplitude and an interchange in the order of deep and shallow minima \citep{Bodi-2016}. The origin of these changes is as of yet unclear, but we note that the change of the order in period-doubled light curves was observed elsewhere, for example in light curves of RR Lyr and RV Tau stars \citep{Molnar-2012,Plachy-2013} and hydrodynamic models of BL Her variables \citep{Smolec-2012}.

Cepheid stars also highlight one of the great advantages of the K2 mission over the primary one. Although the step-and-stare approach limits the baseline of individual observations, targeting multiple directions and different stellar populations can accumulate measurements of rare objects. \textit{Kepler} is expected to observe dozens of Cepheids, from all subgroups, in the Galactic field, and potentially hundreds of stars during the campaigns that target the Galactic bulge. The first results from two W Vir-type stars (medium-period type II Cepheids) has already led to the detection of significant cycle-to-cycle variations that will be undoubtedly followed by other discoveries in the next few years (Plachy et al., in prep).

\section{RR Lyrae stars}
Whereas the \textit{Kepler} data of the Cepheid star were intriguing, the observations of RR Lyrae stars turned out to be downright revolutionary. The discovery of period doubling (alternating low- and high-amplitude cycles) and the new modeling efforts it sparked were already presented in the recent review of \citet{2012JAVSO..40..481K}. Our picture got much more complicated compared to the simple categorization of RRab (fundamental mode), RRc (first overtone) and RRd (double mode) subtypes that we used less than a decade ago. The discovery of low-amplitude additional modes blurred the line between the three main groups, especially for the RRab class \citep{Benko-2010,Benko-2014}. According to space-based photometry, fundamental-mode stars that do not show the Blazhko effect have no additional modes either, i.e., they are true single-mode pulsators. Blazhko stars, on the other hand, often (but not always) pulsate in a few additional modes as well. The prototype, RR Lyrae is also a member of the latter group \citep{Molnar-2012}. Hydrodynamic modes managed to explain some, but not all of these discoveries: period doubling, for example, can destabilize the first overtone in stars that are otherwise away from the double-mode regime of RR Lyrae stars \citep{Molnar-2012a}. What's more, the interplay of three different modes opens the possibility for intricate dynamical states, including low-dymensional chaos in the models \cite{Plachy-2013a}. Let's just think about that for a moment: the observations of \textit{Kepler}, and the modeling work that it sparked, transformed these stars from ``simple" radial pulsators into stars where even chaos may occur. 

First-overtone and double-mode stars also held some surprises. As \textit{Kepler} was busy collecting data, new results from the ground-based OGLE survey indicated that some first-overtone Cepheids and even a few RRc stars share similar additional modes in the $f_1/f \sim$ 0.60--0.64 frequency range \citep{Moskalik-2009,2010AcA....60...17S}. As more and more data poured in from OGLE, these modes started to form three distinct groups, and they were also found in double-mode stars, but still connected to the first overtone. But the precision of the OGLE data, even after a decade of observations, is limited compared to what \textit{Kepler} was able to do, so it was still unclear just how frequent these weird modes are. And although the data from \textit{Kepler} (and the first K2 observations) were limited to a handful of RRc and RRd stars, they suggested an intriguing answer: these modes are there in all RRc and RRd stars \citep{Kurtz-2015,Molnar-2015,Moskalik-2015}. This means that RRc and RRd stars are not single- and double-mode stars anymore, instead they pulsate at least in two and three modes, respectively (not counting any further modes seen is some of them). These additional periodicities most often do not fit the radial (axisymmetric) oscillation spectrum, so the most plausible explanation is the existence of nonradial (non-axisymmetric) modes in these objects.

\begin{figure*}
\includegraphics[width=1.0\textwidth]{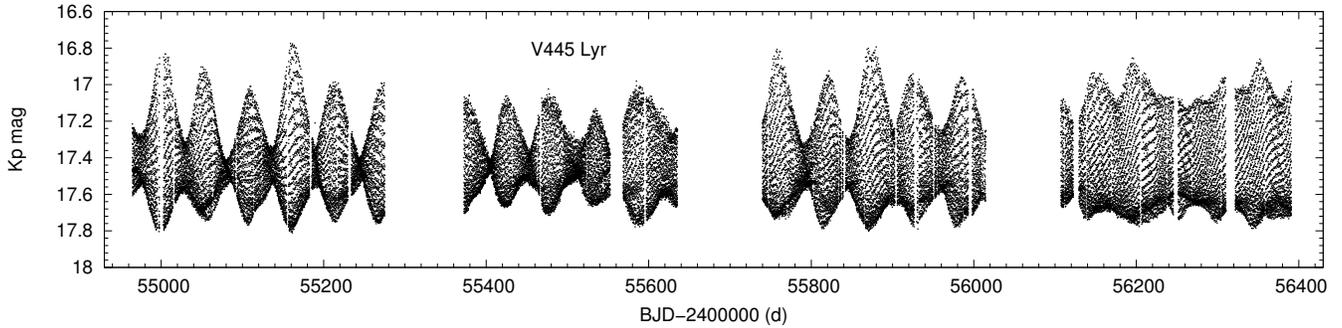}
\caption{The complete \textit{Kepler} light curve of the Blazhko RR Lyrae star V445 Lyr. Smaller gaps are caused by data download and safe mode periods, the three long ones by the failure of one of the CCD modules. Note the very strong modulation at the beginning that decreased to low levels towards the end. }
\label{fig:v445}
\end{figure*}

The benefits of the continuous observations of \textit{Kepler} are clearly illustrated when we look at the RR Lyrae light curves as a whole. Ground-based surveys suggested that some Blazhko stars have multiple or irregular modulation cycles, but the space-based data provided us with clear evidence \citep{LaCluyze-2004,Sodor-2010}. In the original \textit{Kepler} field, 80\% of the modulated stars turned out to be ``complicated", with changing or multiperiodic Blazhko cycles. V445 Lyr, for example, was heralded as one of the most extremely modulated stars observed, with pulsation amplitudes falling from 1.5 mag to 0.15 mag, and the light curve getting completely distorted at minimum amplitude, based on the first few observing quarters \citep{Guggenberger-2012}. Within a few years though, the modulation pattern barely resembled those extreme values, and the amplitude variation decreased from $>1$ mag to a meager 0.3 mag, as Fig.~\ref{fig:v445} illustrates. These observations further highlight that a large portion of RR Lyrae stars are not just simple, repetitive pulsators, but are intricate physical laboratories that hold many more puzzles to solve.

Unfortunately, the K2 observations are limited to a maximum of 80 days per field that is often far too short to cover even a single modulation cycle. Nevertheless, K2 can be exploited for other purposes. For example, the very first engineering run already delivered us the first observations of two RRd and a modulated RRc stars that were lacking from the original field \citep{Molnar-2015}. The large aperture of \textit{Kepler} (especially compared to other photometric missions) also allows us to reach faint stars down to 20-22 mag. Given the intrinsic brightness of RR Lyrae stars (and Cepheids as well), we can thus reach the very edges of the halo of the Milky Way, and even the nearest galaxies in the Local Group. One such nearby object is the tiny, faint dwarf galaxy Leo IV, where only three RR Lyraes are known. \textit{Kepler} observed all of them, and detected the Blazhko effect clearly in one, and possibly in another star out of the three \citep{Molnar-2015a}. This represents the first unambiguous observation of the Blazhko effect beyond the Magellanic Clouds.

\section{Main-sequence pulsators}

Let us now turn back to the Main Sequence, to the young and large stars that populate the HRD above the Sun. Here, many types of variable stars can be found, from the largest O and B-type ones to the smaller $\delta$\,Scuti- and $\gamma$\,Doradus-type pulsators. O-type blue supergiants are very rare, and given their short life times of just a few tens of millions of years, they can be observed where they were formed. The original \textit{Kepler} field lacked any star-forming regions, an thus O stars, but the K2 mission successfully covered a few of them. The initial results are puzzling. Many things seem to happen simultaneously in the light curve, including rotational modulation from chemical and/or temperature spots and various pulsation modes that are hard to disentangle \citep{Buysschaert-2015}. 

At slightly lower temperatures, B stars are well-known pulsators, hosting two groups: the $\beta$\,Cephei-type stars (not to be confused with the $\delta$\,Cephei group), and the SPB, or Slowly Pulsating B stars, pulsating in \textit{p}- and \textit{g}-modes, respectively. However, many stars are \textit{hybrids} that pulsate in both regimes simultaneously. From there, it is just a matter of observing enough modes to carry out an in-depth analysis and modeling of the structure of these stars, analogous to the study of solar-like oscillations \citep{Moravveji-2015,Papics-2015}. And the analysis of the \textit{Kepler} data revealed some intriguing insights into the inner workings of massive stars. 

We instinctively assume, for example, that stars rotate in one direction, with some differences in the actual speed of the rotation. However, B stars tell a different story: it seems that in some stars the direction of rotation turns around at a given radius, and the upper envelope counter-rotates the core and the deeper layers \citep{Triana-2015}. This also means that some layers of the star are essentially at standstill, and have no angular momentum. Our current stellar evolution models cannot explain these observations. 

Moving downwards, the next large group is the $\delta$\,Scuti-$\gamma$\,Doradus variables that have spectral types ranging from late A to early F stars. However, recent studies unearthed some variable stars that fall into the gap between the instability strips of classical SPB and A-type pulsators. These are currently called Maia-type and hot $\gamma$\,Dor stars, although their relation with the other groups and the excitation mechanisms behind their variation are both uncertain \citep{Balona-2015,Balona-2016}. We note that \textit{Kepler} also observed Maia (20 Tau), a member of the Pleiades cluster, during Campaign 4, so the namesake of the group can also be studied.

Turning back to $\delta$\,Scutis: these stars pulsate in radial and nonradial \textit{p}-modes with somewhat shorter periods than $\gamma$\,Dor stars that exhibit \textit{g}-modes. 
On the Hertzsprung-Russell diagram stars belonging to these classes overlap, therefore {\it hybrid stars} that would show both \textit{p}- and \textit{g}-modes are expected in these groups as well. Indeed, from ground-based observations a few such hybrid oscillators have been found. These are valued treasures, since more oscillation modes reveal mode information about the stellar structure, in addition \textit{p}- and \textit{g}-modes probe different domains within the star. Enters \textit{Kepler} which targeted hundreds of stars in this region of the parameter space. The result: too many (over 23\% of) stars show periodic variations in these frequency regimes \citep{Uytterhoeven-2011}. Maybe not all of these objects are genuine hybrids, since there are other possible explanations for the long-period variability: instrumental origin, stellar spots, binarity or ellipsoidal variations. Confirming or refuting these alternative scenarios takes time and follow-up observations with many telescopes, but the hunt for genuine hybrid stars continues, in order to solve the mysteries they present.

Stellar pulsation can be exploited to investigate stars that themselves may not be variables, but are in a binary system with one \citep{Murphy-2015}. The method is based on the idea that if a standard clock is placed on an orbit around a star, the signal from the clock will arrive sooner or later to the observer depending on the position of the clock along the orbit \cite{Shibahashi-2012}. In the studied systems the clock are the independent pulsation frequencies of the variable star component. If periodic O--C modulation, or variations in  period/frequency or in phase can be detected in these stars, then one might conjecture that the star has a companion. It is important that simultaneous variation should be seen in all independent frequencies, and these variations should be in phase in order to exclude any spurious, intrinsic variations of the oscillation modes. And this is not a hypothetical issue: the same \textit{Kepler} observations also revealed that a fraction of modes in $\delta$\,Scuti stars indeed show intrinsic amplitude or phase variations, complicating such studies \citep{Bowman-2016}.

But back to the O--C studies. In practice only the modes with the largest amplitudes can be studied, but this new method---allowed by the long, precise \textit{Kepler} data---is a powerful technique to discover new binary systems. Its advantage is that no eclipses, e.g., lucky geometric configurations are required, binary systems with arbitrary inclinations -- excluding almost face-on configurations -- can be found. The light-time effect, because its base is the finite speed of light, allows hundreds of binary systems to be discovered. And it has an unbeatable advantage: there is no need for time-consuming spectroscopy, still radial velocity curve can be obtained from photometry alone! The astrophysical importance of the method lies in the fact that statistics of companions can be obtained for higher mass stars which are notoriously difficult to study with other techniques. With this method -- which can be considered as a new planet-finding technique -- even a high-mass planet (with 12 Jupiter masses) was discovered around an A-type star on a 840-day orbit \citep{Murphy-2016}, demonstrating the potential and complementary nature of the new high-precision space photometry and the more conventional radial velocity methods.

\section{Pulsating subdwarfs and white dwarfs}

A number of variables called hot subdwarfs can be found in the region between the zero-age main sequence and the white dwarf cooling track, populating the so-called extended horizontal branch. These subdwarf stars are believed to be the exposed cores of stars that should be on the normal horizontal branch, but experienced extreme mass loss and basically shed their envelopes when they were red giants. These objects  show non-radial pulsations in either \textit{p}- or \textit{g}-modes or in both of them. Asteroseismic analysis of these pulsation modes led to strong constraints on the surface properties and inner structures of these stars as well. Thanks to the \textit{Kepler} data, mode identification techniques using the observed period spacings and frequency multiplets became a reality. One interesting discovery about sub-dwarfs is that their rotation periods are surprisingly long (10-100 days) even for short-period (~0.5 days), close binary systems, where tidal forces are expected to synchronize the orbital and rotation periods \citep{Reed-2014}. 

White dwarfs are the remnant products of stellar evolution for the vast majority of stars. As the mission progressed, numerous white dwarfs were discovered in the original \textit{Kepler} field of view. Unfortunately, however, only few were actually selected for observation before the sudden end of the mission \citep{Greiss-2016}. But these \textit{Kepler} measurements led to a discovery of an entirely new phenomenon in pulsating white dwarfs: the occasional large-amplitude outbursts \cite{Bell-2015}. The ~1.5-year long data with one minute sampling of KIC\,4552982, a pulsating white dwarf that belongs to the ZZ\,Cet subclass, shows 178 such events in which the flux of the star increased by several per cent, lasting from a few hours to a day. With the two-reaction-wheel mission the number of known outbursters increased to four by now \citep{Hermes-2015,2016arXiv160701392B}. These discoveries suggest that such outbursts could be a common phenomenon that was either missed, by chance, during the past decades or was simply removed during of the data processing. Ground-based observations are affected by atmospheric extinction that manifests as a slow brightening and fading as a star rises and sets, and it is seen through different amounts of air. These effects are easy to confuse with similar, but intrinsic variations. 

Nevertheless, detailed theoretical explanation of this phenomenon is still needed. The four outbursters are among the longest pulsation period ones located at the red edge on the DAV instability strip, where mode coupling could be prevalent, therefore consideration of pulsation energy transfer via nonlinear resonances is plausible. Such discoveries show the power of space-based, continuous photometry that is free from the biases induced by the rotation period of our own planet. 

\section{Cataclysmic variables}

\begin{figure*}
\includegraphics[width=1.0\textwidth]{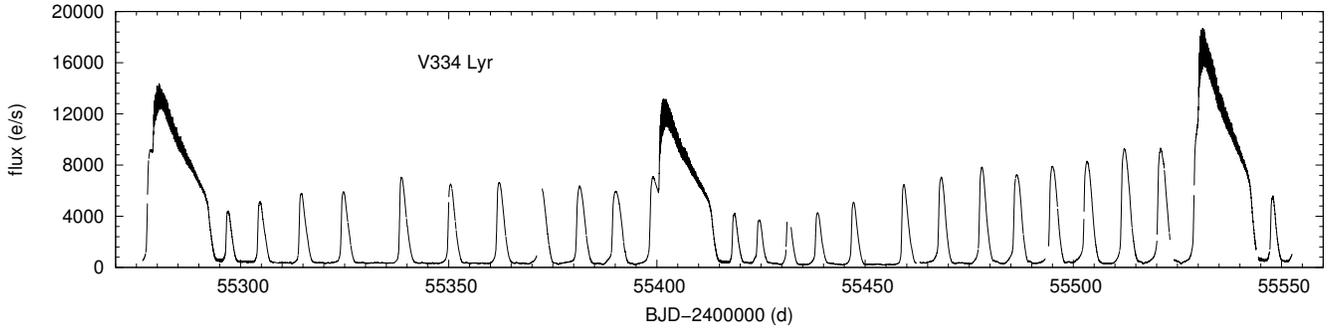}
\caption{Light curve of V334 Lyr, an SU\,UMa-type dwarf nova from \textit{Kepler} quarters 5, 6, and 7, spanning 276 days. Notice the normal and superoutbursts throughout observations. In the case of all superoutbursts a trigger (normal) outburst can be seen.}
\label{fig:v334lyr}
\end{figure*}

Cataclysmic variable binary systems consist of a low-mass main-sequence stars that transfer mass onto a white dwarf primary via an accretion disk. These variables (dwarf and recurrent novae, nova-like objects) are bona-fide amateur objects. Their erratic nature, the unpredictable, large-amplitude outbursts, and the involved timescales between consecutive outbursts (often decades or even centuries) make them ideal targets for a keen amateur observer. So what happens, if one is able to observe them continuously, with exquisite precision and with a large-aperture telescope? This sounds like a dream, and indeed one of the authors (RSz) was oftentimes dreaming about such a possibility while observing in his parents garden some 25 years ago with a 10-inch Dobson-telescope (and submitted his observations to AAVSO, as well). \textit{Kepler} offered just that and observed at least 16 cataclysmic variables in the original \textit{Kepler} field continuously, with unprecedented precision, revealing subtle details and without missing a single outburst during the 4-year mission in most of the cases. The sample contains 6 SU\,UMa type variables, 1 of the WZ\,Sge type, 2 nova-like variable stars, 3 UG-type cataclysmic variables, 2 additional UG-types showing eclipses, and 1 AM\,Her variable. Even new cataclysmic variables were discovered, such as the nova-like, potentially SW\,Sex-type KIC\,8751494 \citep{Williams-2010}, or a background SU\,UMa-type variable in the photometric aperture of a G-type exoplanet target star \citep{Barclay-2012}.

For example, early \textit{Kepler} data revealed superhumps from V334\,Lyr (Fig.~\ref{fig:v334lyr}) during quiescence, normal outburst, and the longer and brighter superoutburst \citep{Still-2010}, whereas so far these variations were detected only during superoutbursts. The orbital period of V334\,Lyr is 2.11\,hr, while the system displays both positive and negative superhumps with periods of 2.20 and 2.06\,hr (longer and shorter than the orbital period), respectively. The quality of the \textit{Kepler} data is such that it is possible to test the models for accretion disk dynamics quantitatively. The  V344\,Lyr data are consistent with the model that two physical sources yield positive superhumps: early in the superoutburst, the superhump signal is generated by viscous dissipation within the periodically flexing disk. But late in the superoutburst, the signal is generated as the accretion stream bright spot sweeps around the rim of the non-axisymmetric disk. The V344\,Lyr data also reveal negative superhumps arising from accretion onto a tilted disk precessing in the retrograde direction. It is also found that at least some long superoutbursts appear to be triggered by short outbursts in some SU\,UMa stars (see, e.g., Fig.~\ref{fig:v334lyr}).

\section{Supernovae}

Supernova explosions are unexpected events, so it is very hard to catch the first hours of the explosion that may hold vital information about their progenitors. The continuous monitoring of extragalaxies provide a great possibility to observe these events from the very beginning and not only when the brightening is already in full swing. Type Ia supernovae are crucial objects in extragalactic distance measurements, therefore the importance of understanding the mechanisms driving the thermonuclear runaway reaction is unquestionable, so the targets of \textit{Kepler} included some galaxies as well in its Guest Observer program, where all interested scientist could propose targets in \textit{Kepler}'s field of view. Overall, three Type Ia supernovae were observed in the original \textit{Kepler} mission. No signatures of interaction between the ejected material and potential companion stars were found, suggesting that the progenitor systems contained two compact stars and the explosion was caused by their merger \citep{Olling-2015}. However, we cannot simply replace the classical, accretion-induced collapse theory with the double-degenerate scenario. Recent ground-based observations found evidence for a non-degenerate binary companion to a normal SN Ia \citep{2016ApJ...820...92M}. According to theoretical considerations both channels contribute to the observed SNIa events, but their relative importance is largely unknown. Early space-based photometric supernova observations in a significant number of cases may answer these question in the future and inform us about the progenitors and their circumstellar environments in the other types of supernovae, too. 

The K2 mission observed a few fields with large numbers of galaxies, but it have not discovered as many supernovae as it was hoped yet. Nevertheless, it already measured two transient events that were identified as core-collapse supernovae \cite{Garnavich-2016}. Thanks to \textit{Kepler}, we were able to detect the shock breakout for the first time in the optical band. This short, sudden brightening precedes the actual explosion: it happens when the outwards shockwave from the collapsing core breaches the surface of the star. To give a sense of scale here: after the core starts to collapse, it takes about one day for the shockwave to travel to the surface of the star. The star then suddenly (within a few minutes) flares up, and then starts to fade back for a few hours, before the actual disruption begins. From there, it still takes about 10 days for the exploding star to reach peak brightness. 

Interestingly, the shock breakout was only seen in the data of the larger supergiant, while the smaller one showed a slower excess brightening on the ascending branch instead. Both events had similar explosion energies, therefore the diversity of the two rising light curves is puzzling. A possible explanation is that the light from the breakout was absorbed by thicker circumstellar matter in the second case. The circumstellar matter reprocessed that light into heat, and that heating led to a slower but stronger excess brightening. Hopefully the K2 campaigns aimed at nearby clusters will provide us with even more supernova light curves: Campaign 10, for example, includes nearly 5000 selected galaxies in the direction of the Virgo galaxy cluster. 

\section{Eclipsing binaries}

Currently there are 2878 known eclipsing binaries in the original \textit{Kepler} field \citep{kirk-2016}. They come in all flavors (detached, semi-detached, contact, Algol-type, W\,UMa, binaries with third components \citep{Borkovits-2016}), all kinds of strange animals in the eclipsing binary zoo. It is important to know the occurrence rates of different types to identify false positives in the transiting exoplanet search. 
Besides this practical usefulness, there has been a lot of interesting discoveries, where rare types of eclipsing binaries have been discovered. First of all, a long-awaited holy grail was found: a planet orbiting a binary star system. That was Kepler-16 \citep{Doyle-2011}, which is a Saturn-size planet orbiting with a 289-day period around an eclipsing binary. The orbit of the planet is much wider than that of the binary, so it is much like Tatooine (also the unofficial name of the object), which was later found to be quite common in the Milky Way galaxy. 

\begin{figure}
\includegraphics[width=1.0\columnwidth]{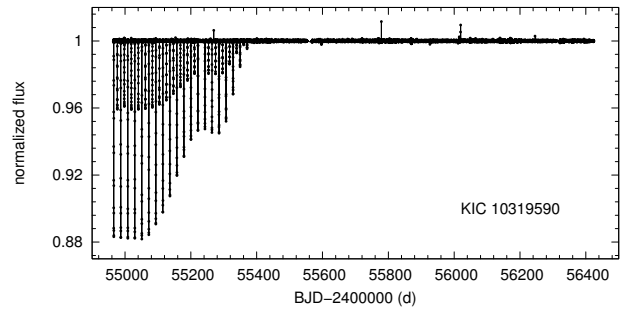}
\caption{Four-year-long \textit{Kepler} light curve of KIC 10319590. The object ceased to be an eclipsing binary.}
\label{fig:103}
\end{figure}

Another one is binaries that cease to show eclipses due to the precession of the plane of their orbit (Fig.~\ref{fig:103} shows an example from the \textit{Kepler} original four-year mission). Although other cases were recorded, such as SS Lac \citep{Zakirov-1990} and HS Hya \citep{Zasche-2012}, \textit{Kepler} was the first where this process could be followed continuously, and the orbital evolution happened in front of our eyes (or in front of our detectors to be more precise).

Yet another feat was the discovery of the triply eclipsing hierarchical triple star, HD\,181068 (dubbed as Trinity by the discoverers \citep{Derekas-2011}). Hierarchical triple systems comprise a close binary and a more distant component. They are important for testing theories of star formation and of stellar evolution in the presence of nearby companions. \textit{Kepler} photometry of Trinity supplemented by ground-based spectroscopy and interferometry showed it to be a hierarchical triple with two types of mutual eclipses. The primary is a red giant that is on a 45-day orbit with a pair of red dwarfs in a close 0.9-day orbit. The red giant shows evidence for tidally induced oscillations that are driven by the orbital motion of the close pair. Thus, HD\,181068 turned out to be an ideal target for studies of dynamical evolution and for testing tidal friction theories in hierarchical triple systems.

Besides stochastically excited solar-like oscillation and the good old $\kappa$-mechanism that drives the classical pulsators from RR Lyrae and Cepheid stars to $\delta$\,Scutis and Miras, there is a third excitation mechanism, which is rarely mentioned. This is tidally-driven oscillations, a process that takes place in certain binary stars. In some cases the components of binary systems get close to each other during their orbital motion, where the gravitational interaction can excite modes that would be damped otherwise (one needs considerable eccentricity for a variable driving force). There is even an emerging new research field, called \textit{tidal asteroseismology}, that benefits from the unusual geometrical configuration that makes us the favor to sound our stars that otherwise would be too shy to oscillate. 

\begin{figure}
  \centering
  \includegraphics[height=1.0\columnwidth, angle=270]{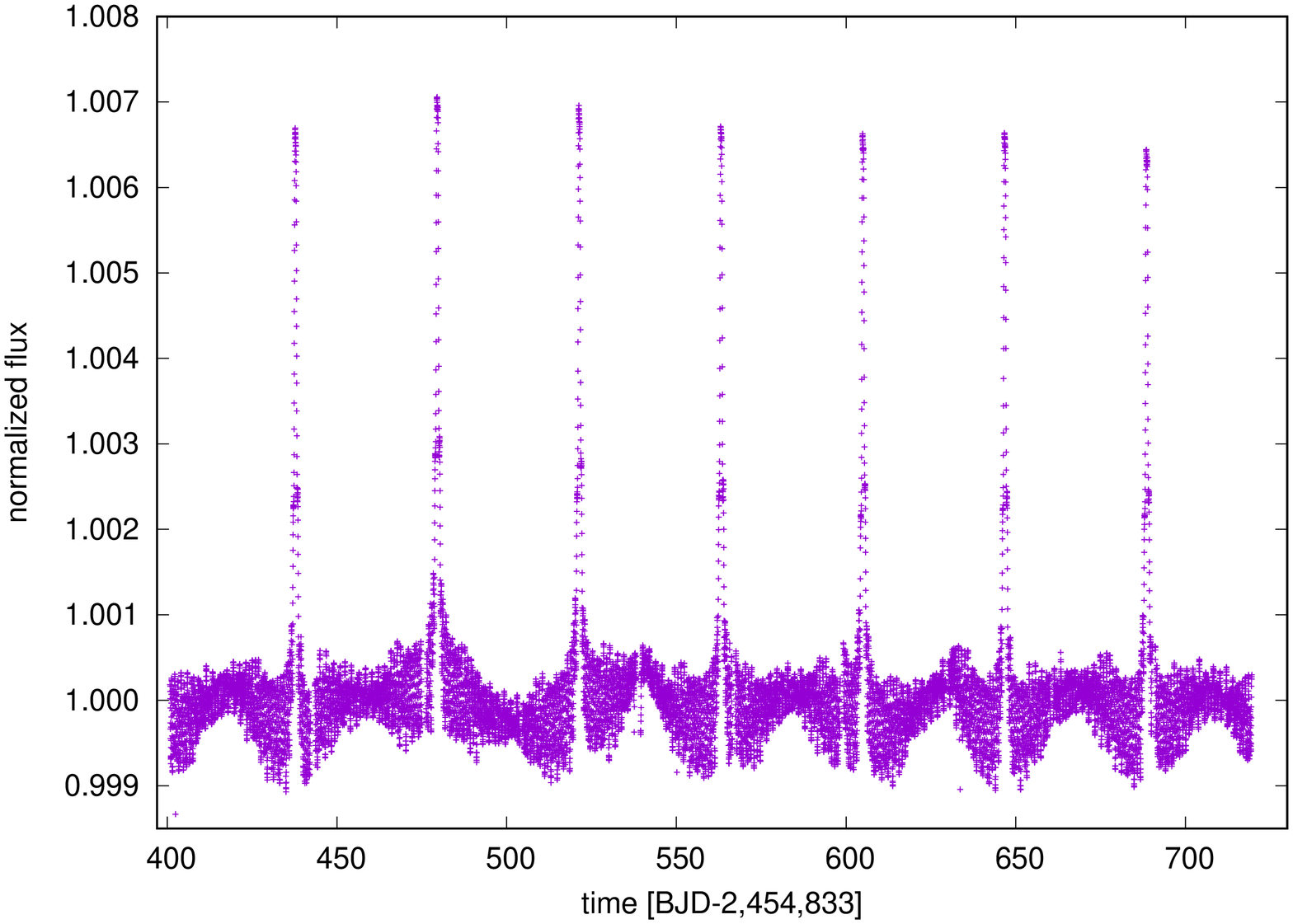}
  \includegraphics[height=1.0\columnwidth, angle=270]{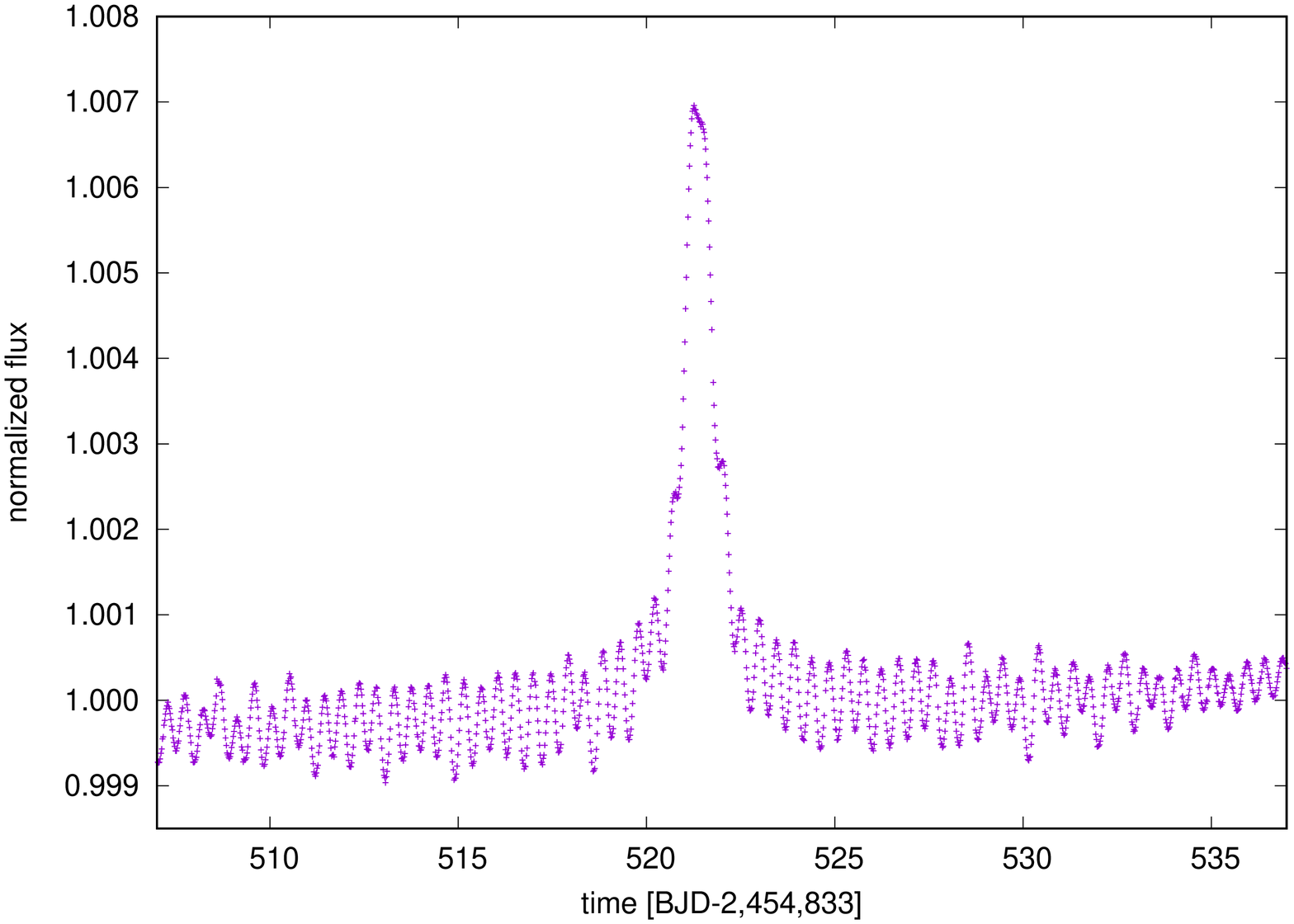}
\caption{Top: The Kepler light curve of KOI-54, a heart-beat star. Bottom: A zoom-in in the light curve of KOI-54. In between the brightenings the system shows tidally induced oscillations.}
\label{fig:koi54}
\end{figure}

A prime example is KOI-54 (aka KIC\,8112039) \citep{Welsh-2011} that made astronomers scratch their heads once it was found in the sea of unbelievable \textit{Kepler} data. The light curve shows distinct, periodic brightenings of 0.7\%, and ubiquitous oscillations in between (see the upper panel of Fig.~\ref{fig:koi54}). First it was thought to be the manifestation of a very exotic binary system (maybe containing a black hole that causes gravitational lensing, hence brightening). Follow-up observations, however, showed something else: a highly eccentric pair of A-type stars dancing around each other on a 46-day orbit and approaching each other within a few stellar radii. The strong gravitational tug deforms the stars at the close encounter, and heats up the stellar hemispheres that face towards each other. This is the cause of the brightening (see the lower panel of Fig.~\ref{fig:koi54} for a zoom-in.). In addition, the gravitational interaction excites oscillations in both components allowing a detailed seismological investigation of otherwise non-oscillating main-sequence stars. The light curve reminded the astronomers an electrocardiogram, so this type of objects got an expressive name: {\it heart-beat stars}. It was soon found out that the system is not unique (in fact the existence of this configuration and the observable features were predicted two decades ago), and many other systems were unearthed from the \textit{Kepler} data \citep{Thompson-2012}.

\section{What's left for amateurs?}

With the influx of new discoveries based on ever more precise data, one could not help but to ask: is there still place for amateur observations, or, indeed, smaller professional telescopes? Perhaps surprisingly, our answer is a wholehearted yes. The abrupt end of the primary mission of \textit{Kepler} left astronomers with literally thousands (if not tens of thousands) of targets in need of follow-up observations. The best-known example is KIC 8462852, or Boyajian's star, the mysterious object that showed irregular, short and deep fadings, most of which occurred during the last 100 days of the 1600-day long mission \citep{Boyajian-2016}. The star is being followed by various amateur and professional astronomers to catch another large ``dip", and after a successful fundraising campaign, the authors also secured extended telescope time on the Las Cumbres Observatory Global Telescope Network (LCOGT) to search for smaller ones.

Turning back to the pulsating stars: one expectation of the \textit{Kepler} mission was to see whether the modulation of the star RR Lyr indeed goes through 4-year cycles, as suggested by \citet{1973IBVS..764....1D}. \textit{Kepler} did not detect any signs of such cycle or an associated shift in the phase of the Blazhko effect: instead the amplitude of the modulation was steadily decreasing towards the end of the primary mission. But where was it heading? Did the Blazhko effect turn off completely? Did it recover? The answer was provided by the observations of amateur astronomers from the AAVSO, compounded by an array of small telephoto lenses called the Very Tiny Telescopes (VTTs). These data allowed the researchers to follow the phase variations of RR Lyr, and revealed that the modulation effectively turned off by 2014, before restarting in 2015 \citep{LeBorgne-2014,2016CoKon.105...73P}. Several papers also used the AAVSO measurements to put the \textit{Kepler} observations into broader temporal context. A few examples include works on Miras and semiregular stars (\citealt{Banyai-2013}, see Fig.~\ref{fig:tucyg}), dwarf novae \citep{Otulakowska-Hypka-2016}, and intermediate polar stars \citep{2016arXiv160901026L}.

Despite the advent of large sky-surveys, like Pan-STARRS, Gaia, and LSST, long-term behavior of pulsating stars, e.g., O--C variations can be secured only through amateur observations (often extending to several decades or even centuries). Perhaps counterintuitively, the largest surveys will not replace the backyard astronomers at all. There is a simple reason for this: the large-aperture telescopes collecting data for Pan-STARRS or LSST are aimed at the billions of faint objects on the sky, but typically saturate in the 12-15 magnitude range, leaving the bright objects unchecked. There are some professional projects that aim to fill this gap, like the Fly's Eye, Evryscope or MASCARA instruments \citep{Pal-2013,Law-2015,Snellen-2013}, but unlike the observing network of the AAVSO, these depend on the funding flow instead of the supply of volunteer observers. The sky is still there to explore and to discover for everybody. 

\section*{Acknowledgements}
The authors have been supported by the Lend\"ulet 2009 and LP2014-17 Young Researchers Programs, the Hungarian National Research, Development and Innovation Office (NKFIH) grants K-115709, PD-116175 and PD-121203, the European Community's Seventh Framework Programme (FP7/2007-2013) under grant agreement no. 312844 (SPACEINN), and the ESA PECS Contract Nos. 4000110889/14/NL/NDe. L.~M. was supported by the J\'anos Bolyai Research Scholarship. The authors wish to thank Dr.~T.~Borkovits for his useful comments on the precession of eclipsing binary systems. This paper includes data collected by the \textit{Kepler} mission. Funding for the \textit{Kepler} and K2 missions are provided by the NASA Science Mission Directorate. This research has made use of the NASA Exoplanet Archive, which is operated by the California Institute of Technology, under contract with NASA under the Exoplanet Exploration Program. The authors gratefully acknowledge Ball Aerospace, the \textit{Kepler} Project, Science, and Guest Observer Offices, and all those who have contributed to the mission for their tireless efforts which have made these results possible.

%% End of file `sample.tex'.

\end{document}